\documentclass[12pt]{article}
\usepackage{graphicx}
\usepackage{amsmath}
\usepackage{amssymb}
\usepackage{caption2}
\begin{document}
\bibliographystyle{prsty}
\begin{center}
{\large {\bf \sc{   The strong coupling constant $g_{D^*  D \pi}$  and final-state interactions }}} \\[2mm]
Zhi-Gang Wang \footnote{E-mail,wangzgyiti@yahoo.com.cn;wangzg@yahoo.cn.  }    \\
Department of Physics, North China Electric Power University,
Baoding 071003, P. R. China
\end{center}

\begin{abstract}
In this article, we study the contribution from the final-state
interactions to the strong coupling constant $g_{D^* D \pi}$. We
take an assumption that the momentum transfers in the strong decay
$D^{*+} \to D^0 \pi^+$ be large  to validate the operator product
expansion in the light-cone QCD sum rules. At large momentum
transfers, the final-state interactions play an important role, and
we should take them into account.
\end{abstract}


\section{Introduction}

Several QCD sum rules  approaches have been applied to determine the
strong coupling constant $g_{D^* D \pi}$ in the strong decay $D^{*+}
\to D^0 \pi^+$, such as two-point correlation function  with
soft-pion technique \cite{SoftP1,SoftP2}, or beyond the soft-pion
approximation \cite{SoftP3}, light-cone QCD sum rules
\cite{Belyaev94,Colangelo98}, light-cone sum rules with perturbative
$\alpha_s$ corrections \cite{Khodjamirian99}, QCD sum rules in a
external field \cite{Grozin98}, double-moment   QCD sum rules
\cite{Dosch96}, and double Borel sum rules \cite{Navarra00}, etc.
The discrepancy between the experimental data from the CLEO
collaboration and the predictions from the QCD sum rules is very
large.
 The upper bound $g_{D^\ast D \pi}=13.5 $ ($g_{D^\ast D \pi}=10.5 \pm 3.0$ from the light-cone QCD sum rules
 with perturbative
 $\alpha_s$ corrections to the twist-2 light-cone distribution amplitude $\phi_\pi(\mu,u)$ \cite{Khodjamirian99}) is too small to
  account for the experimental data, $g_{D^\ast D \pi}=17.9 \pm 0.3 \pm
1.9$ \cite{CLEO}.

It has been noted that the simple quark-hadron duality ansatz which
works in the one-variable dispersion relation might be too crude for
the double dispersion relation \cite{KhodjamirianConf}. In
Ref.\cite{Becirevic03}, the authors  observe  that inclusion of the
contributions from an explicit radial excitation to the hadronic
spectral density can  improve the value of $g_{D^* D \pi}$
significantly, however, additional (strong) assumptions about the
strong coupling constants concerning the radial excitations are
taken. On the other hand, in Ref.\cite{Navarra00}, the authors find
that in the standard QCD sum rules, a modification of the
contribution from the continuum states may lead to unstable sum
rules. In Ref.\cite{Kim03Thr}, the authors argue that the
subtracting term $M^2e^{-\frac{s_0}{M^2}}$ comes from a
mathematically spurious term and it should not be a part of the
final sum rules, however, absence of the continuum states
subtraction seems rather strange.

In Ref.\cite{Duraes04}, the  form-factor $g_{D^* D \pi}(Q^2)$ for
off-shell $D$ meson is evaluated at low and moderate $Q^2$ in a
hadronic loop model. The authors fix the arbitrary constants  to
match previous QCD sum rule calculations valid at higher $Q^2$, then
 extrapolate to the mass shell to obtain the coupling constant
$g_{D^* D \pi}$.

Despite   large uncertainties, the QCD sum rules have given a great
deal of  good agreements with the experiment data. The strong
coupling constant $g_{D^* D \pi}$   seems to be  exotic. In the
heavy quark limit, a quark model  based on the Dirac equation in a
central potential leads to the value $g_{D^* D \pi} \approx 18$
\cite{Dirac99}.  The quenched lattice QCD calculation  results in
$g_{D^* D \pi}=18.8 \pm 2.3^{+1.1}_{-2.0} $ \cite{Latt02}.

We study the strong coupling constant $g_{D^*D\pi}$ with the
two-point correlation function $\Pi_{\mu}(p,q)$
\cite{Belyaev94,Khodjamirian99},

\begin{eqnarray}
\Pi_{\mu }(p,q)&=&i \int d^4x \, e^{-i q \cdot x} \, \langle 0
|T\left\{J_\mu(0) J_5^+(x)\right\}|\pi(p)\rangle \, ,
\\
J_\mu(x)&=&{\bar u}(x)\gamma_\mu   c(x)\, ,  \nonumber \\
J_5(x)&=&{\bar d}(x) i \gamma_5  c(x)\, ,
\end{eqnarray}
where the currents $J_\mu(x)$  and $J_5(x)$ interpolate the mesons
$D^{*}$ and $D$, respectively.  The external state $\pi$ has the
four momentum $p_\mu$ with $p^2=m_\pi^2$. In this article, we take
 the isospin limit for the $u$ and $d$ quarks.

The calculations are performed at  large spacelike momentum regions
$(q+p)^2\ll 0$ and $q^2\ll 0$, which correspond to small light-cone
distance $x^2\approx 0$ required by validity of the operator product
expansion \cite{LCSR,LCSRreview}. In the strong decay $D^{*+} \to
D^0\pi^+$, the momentum transfers $D^* \to \pi$ and $D^* \to D$ are
very small, $m_{D^*}\approx m_D+m_{\pi}=(1.87+0.14)\rm{GeV}$. In the
light-cone QCD sum rules, we perform the operator product expansion
at large momentum transfers, at that energy scale, the final-states
  are active, their interactions may play an important
role and we should take them into account.   In this article, we
study the final-state interactions (elastic scatterings) of  $ D\pi$
with   Bethe-Salpeter re-summation.

Take the amplitudes from the chiral Lagrangian as kernels, and solve
the corresponding Bethe-Salpeter equation, we can re-sum an infinite
series of loop diagrams in chiral expansions,     generate
quasi-bound states of the mesons (or baryons) dynamically, and
account for the resonances without including them explicitly
\cite{BSE}.

The article is arranged as: in Section 2, we perform Bethe-Salpeter
re-summation for the final-state interactions in the strong  decay
$D^{*+} \to D^0 \pi^+$; in Section 3, the numerical result and
discussion; and in Section 4, conclusion.

\section{Bethe-Salpeter re-summation for the final-state interactions }
In order to describe the interactions between the light and  heavy
pseudoscalar mesons, we employ the leading order heavy chiral
Lagrangian \cite{HeavyChiral},
\begin{equation}
 {\cal L} =
\frac{1}{4f_{\pi}^2}\left\{\partial^{\mu}P[\Phi,\partial_{\mu}\Phi]P^{\dag}
- P[\Phi,\partial_{\mu}\Phi]\partial^{\mu}P^{\dag}\right\},
\end{equation}
where $f_{\pi}=92.4\rm{MeV}$  is the weak decay constant of the
$\pi$, $P$ stand for the charmed mesons  $D^0$, $D^+$ and $D_s^+$,
and $\Phi$ denote the octet pseudoscalar mesons,
\begin{equation}
 \Phi = \left(
\begin{array}{ccc}
\frac{1}{\sqrt{2}}\pi^0 + \frac{1}{\sqrt{6}}\eta & \pi^+ & K^+\\
\pi^- & -\frac{1}{\sqrt{2}}\pi^0 + \frac{1}{\sqrt{6}}\eta & K^0\\
K^- & \bar{K}^0 & - \frac{2}{\sqrt{6}}\eta
\end{array}
\right).
\end{equation}
  The amplitude $V$ for the elastic scattering  $D^+\pi^-\rightarrow D^+\pi^-$
can be obtained from the leading order heavy chiral Lagrangian,
\begin{equation}
 V_{D^+\pi^-}(s,t,u) =
\frac{s-u}{4f_{\pi}^2} \, ,
\end{equation}
where the $s$, $t$ and $u$ are Mandelstam variables \footnote{For
technical details, one can consult
 the Ph.D thesis (in Chinese) of F. K. Guo, Institute of high energy physics.},
\begin{eqnarray}
-u(s,\cos{\theta})&=& s-m_{\pi}^2-m_D^2 -
2\left(m_D^2+\frac{\lambda^2(s,m_\pi^2,m_D^2)}{4s}\right)
   +\frac{\lambda^2(s,m_\pi^2,m_D^2)}{2s} \cos{\theta}, \nonumber\\
 \lambda(s,m_D^2,m_\pi^2)&=&\sqrt{[s-(m_D+m_\pi)^2][s-(m_D-m_\pi)^2]} \, .
\end{eqnarray}
In unitary chiral perturbation theory, with on-shell approximation,
the full scattering amplitude $T$ can be converted into an algebraic
Bethe-Salpeter equation \cite{BSE},
\begin{eqnarray}
T&=&(1-VG)^{-1}V=V+VGV+VGVGV+\cdots ,
\end{eqnarray}
where $V=V_{D^+\pi^-}(s,t,u)$ and
\begin{eqnarray}
G(p^2)&=&i\int\frac{d^4q}{(2\pi)^4}\frac{1}{q^2-m_D^2+i \epsilon}
\frac{1}{(p-q)^2-m_\pi^2+i\epsilon}\,, \\
\text{Re}G(s)&=&\frac{1}{4\pi^2}{\text P}\int_0^{q_{max}} dq \frac{{
q}^2(\omega_D+\omega_\pi)}{\omega_D\omega_\pi
[s-(\omega_D+\omega_\pi)^2]}\, , \\
\text{Im}G(s)&=&-\frac{\lambda(s,m_D^2,m_\pi^2)}{16\pi s} \,,
\end{eqnarray}
here  ${\text P}\int $ stands for the principal integral,
$\omega_i=\sqrt{{ q}^2+m_i^2}$, $q=|\overrightarrow{q}|$ in Eq.(9).

Taking into account the final-state interactions of the $D$ and
$\pi$, the strong coupling constant takes the following form,
\begin{eqnarray}
g_{D^*D\pi}\rightarrow
gg_{D^*D\pi}=g_{D^*D\pi}-g_{D^*D\pi}G(s)T_1(s) \, ,
\end{eqnarray}
here we introduce $g$ to denote the enhanced form-factor comes from
the final-state interactions.
\begin{eqnarray}
T(s,t,u)&=&\sum_{l=0}^\infty (2l+1)L_l(cos\theta)T_l(s) \, ,
\end{eqnarray}
where $L_l(cos\theta)$ are Legendre polynomials, the strong decay
$D^{*+} \to D^0\pi^+$ takes place through relative $P$-wave, we take
the $l=1$ partial wave amplitude $T_1(s)$.

\section{Numerical result and discussion}

There is a singular point at
\begin{eqnarray}
s-(\sqrt{q^2+m_D^2}+\sqrt{q^2+m_\pi^2})^2=0 \, ,
\end{eqnarray}
in the principal integral. In this article, we take the value of the
$q_{max}$  be the typical energy scale $q_{max}=m_\rho=0.77\rm{GeV}$
in the chiral perturbation theory, the value of $s$ should be $s>7.9
\rm{GeV}^2$ to avoid the singular point. If we take  $s$ be the
center of mass of the vector meson $D^*$, $s=m_{D^*}^2$, then
$q_{max}\approx 0$, the final-state interactions  are of minor
importance and can be neglected safely.

In Fig.1, we plot the   enhanced  factor $|g|$ with the variation of
the center of mass parameter $s$. From the figure, we can see the
value of $|g|$ increases quickly according to   $s$.

We take the momentum transfer $D^* \to \pi$  in the strong decay
$D^{*+} \to D^0 \pi^+$  be large  to validate the operator product
expansion in the light-cone QCD sum rules. The deviation
$s-m_{D^*}^2$ measures the virtuality of the initial vector meson
$D^*$, we introduce the effective mass $m_{eff}$ to denote the
virtual mass of the vector meson $D^*$, where the momentum transfer
in $D^* \to \pi$ is large enough.  If we choose the typical value
$s=m^2_{eff}=(m_{D^*}+m_D)^2$, the enhanced factor $|g|$ is rather
large, $|g|\approx 1.4 $. We can take the value from the light-cone
QCD sum rules with perturbative $\alpha_s$ corrections as input
parameter, $g_{D^*D\pi}=10.5\pm 3.0$ \cite{Khodjamirian99},  the
value $|g|g_{D^*D\pi}=14.7\pm 4.2$ is compatible with the
experimental data $g_{D^\ast D \pi}=17.9 \pm 0.3 \pm 1.9$
\cite{CLEO}.

\begin{figure}
\centering
  \includegraphics[totalheight=8cm,width=10cm]{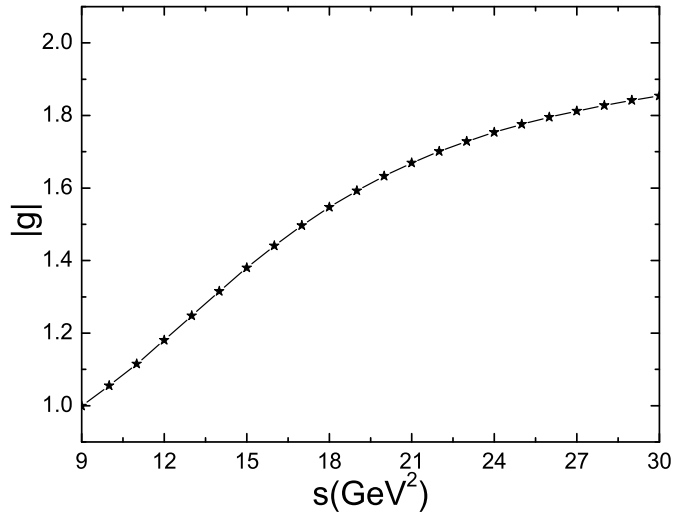}
           \caption{   $|g|=|1-G(s)T_1(s)|$  with     $s$. }
\end{figure}

Although the value of the effective mass $m_{eff}$ suffers from
large uncertainty, we can expect the final-state interactions
 improve the value of the strong coupling constant
$g_{D^*D\pi}$ significantly.  Furthermore, the values of  the strong
coupling constants $g_{D^*D^*P}$, $g_{D^*DP}$, $f_{D^*DV}$,
$f_{D^*D^*V}$, $g_{DDV}$,  $g_{D^*D^*V}$ and $g_{\Delta N \pi}$ from
the light-cone QCD sum rules are  much smaller than most of the
existing estimations or experimental data \cite{Wang07,Wang07B}.
That maybe a general feature of the light-cone QCD sum rules. We
perform the operator product expansion at large momentum transfers,
experimentally, the momentum transfers in the strong decays do not
always  warrant validity of the operator product expansion in the
light-cone. If we take an assumption that the momentum transfers are
large enough, we should take into account all the quantum effects,
because at that energy scale, the final-states are active, their
interactions may play an important role.

\section{Conclusion}
In this article, we study the final-state interactions in the strong
decay $D^{*+}\to D^0 \pi^+$. We take the momentum transfer $D^* \to
\pi$
 be large   to validate the operator product expansion in the
light-cone QCD sum rules. At large momentum transfers, the
final-state interactions play an important role, and we should take
them into account. Although the value of the effective mass
$m_{eff}$ of the vector meson $D^*$ suffers from large uncertainty,
we can expect the final-state interactions   improve the value
significantly .

\section*{Acknowledgments}
This  work is supported by National Natural Science Foundation,
Grant Number 10405009,  and Key Program Foundation of NCEPU.

\end{document}